\documentclass[11pt]{article}
\usepackage{amsmath,amsthm,amssymb,color,graphicx}
\usepackage{braket}

\makeatletter
\def\@normalize{@setsize\normalsize{12pt}\xpt\@xpt
\abovedisplayskip 10pt pluse2pt minus5pt\belowdisplayskip
\abovedisplayskip \abovedisplayshortskip \z@ plus3pt\belowdisplayshortskip
6pt plus3pt minus3pt\let\@listi\@listI}
\def\subsize{\@setsize\subsize{12pt}\xipt\@xipt}
\def\section{\@startsection{section}{1}{\z@}{8pt plus 2 pt minus 2 pt} {8pt plus 2pt
minus 2pt}{\normalsize\bf}}
\def\subsection{\@startsection{subsection}{2}{\z@}{12pt plus 2 pt minus 2 pt}{12pt plus
2pt minus 2pt}{\subsize\bf}}
\makeatother
\begin{document}
\date{}
\title{\large\bf Periodic Orbit Theory Revisited in the Anisotropic Kepler Problem}
\author{\large Kazuhiro Kubo$^1$ and Tokuzo Shimada$^2$\\[5mm]
{\small 
\it{$^1$Max-Planck-Institut f\"ur Physik Komplexer Systeme,}}\\ 
{\small 
\it{
N\"othnitzer Stra\ss e 38, 01187 Dresden, Germany}}\\[3mm]
{\small 
\it{
$^2$Department of physics, School of Science and Technology, 
Meiji University,}}\\
{\small 
\it{
Higashimita 1-1-1, Tama, Kawasaki, 
Kanagawa, 214-8571, Japan
}}\\[2mm]
{\small \it E-mail: kazuhiro@pks.mpg.de, tshimada@gravity.mind.meiji.ac.jp}
}
\maketitle

\begin{abstract}%
The Gutzwiller's trace formula for the anisotropic Kepler problem is Fourier transformed 
with a convenient variable $u=1/\sqrt{-2E}$ which takes care of the scaling property of the 
AKP action $S(E)$. Proper symmetrization procedure (Gutzwiller's prescription) 
is taken by the introduction of half-orbits that close under symmetry transformations
so that the two dimensional semi-classical formulas match correctly the quantum subsectors
$m^\pi=0^+$ and $m^\pi=0^-$. 
Response functions constructed from half orbits in the POT side are explicitly given. 
In particular the response function $g_X$ from $X$-symmetric half orbit 
has an amplitude where the root of the monodromy determinant is inverse hyperbolic. 

The resultant {\it weighted densities of periodic orbits $D^{m=0}_e(\phi)$ and $D^{m=0}_o(\phi)$} 
from both quantum subsectors show peaks at the actions of the periodic orbits 
with correct peak heights corresponding to the Lyapunov exponents of them.  
The formulation takes care of the cut off of the energy levels,
and the agreement between the $D(\phi)$s of QM and POT sides is observed 
independent for the choice of the cut off.
The systematics appeared in the densities of the periodic orbits 
is explained in terms of features of periodic orbits. 

It is shown that from quantum energy levels 
one can extract the information of AKP periodic orbits, even the Lyapunov exponents-- 
the success of inverse quantum chaology in AKP. 
\end{abstract}

\maketitle
\vspace{0.5cm}
\section{Introduction}
\subsection{Quantum-classical correspondence and quantum chaos}
There is a salient distinction in quantum-classical correspondence between the two cases, 
namely, the case of classical theory being integrable and the case of that being chaotic.
In the former, 
the number ($M$) of constants of motion is equal to 
the number ($N$) of  the degrees of freedom (i.e. $M=N$) and 
the classical system trajectory is confined on a $N$-dimensional torus.
Then, the EBK condition 
gives a semi-classical quantization scheme
 \cite{einstein17,brillouin26,keller58}. 
In the latter case ($M < N$), on the other hand, the trajectories are 
exponentially sensitive to the initial conditions. A chaotic trajectory 
covers the constant-energy surface in phase space uniformly and
the phase-space average of some quantity will be equal to
the time-average along any one of the trajectories, if the system 
is entirely in the ergodic regime. 
Because of the absence of the torus, the EBK quantization is no longer applicable
\footnote{Einstein gives a coordinate-independent topological condition
$
\int_{C_i} \vec{p}\cdot d\vec{q} =n_i h~(i=1, \cdots, N)
$.
The Keller's improvement accounts 
for the leakage of the wave function over the turning points 
and gives a condition on the wave number $k$
\begin{align*}
k \oint_{C_i}
\vec{\nabla} S \cdot d\vec{q}=2\pi n_i
+\pi/2~m_i~(i=1, \cdots, N), 
\end{align*}
where 
$m_i$ is the number of the turning points.  This may be deemed as 
a higher dimensional extension of WKB approximation. 
It is interesting to note \cite{dstone05} that Einstein 
already noticed that his quantization condition cannot be applied to the 
classically chaotic case as early as in 1917 \cite{einstein17}.
For the mixed dynamics, the EBK condition can be used only for the 
integrable part.
}.
The {\it quantum chaos} (QC) is the research field which aims 
at the elucidation of quantum-classical correspondence for this latter case.
We call below a quantum system whose classical counterpart is chaotic
as a {\it QC system} for short.

The quantum system (in the non-relativistic regime) is described by the
Schr\"odinger equation which is a linear differential equation for the wave function
of the system. This by no means denies the quantum signature of chaos. 
In fact, with the advent of nanophysics techniques, the  quest for it has become 
tractable in modern experiments. 
We refer to the following pioneering works  for various activities;
conductance fluctuations in quantum dot \cite{marcus92}, 
magnetoresistance on a superlattice of antidots \cite{weiss91}, 
a cold-atom realization of the kicked top \cite{chaudhury09},
wave dynamics \cite{stein92} for the scars \cite{heller84,heller89}, 
the anisotropic Kepler problem (AKP: the topic of the present paper) 
\cite{zhou09,zhou10},
and the topological insulators \cite{hasan10} may shed 
light on tenfold way (see shortly below) and the {\it periodic table} based on Bott periodicity 
\cite{kitaev09, ryu10}.

There are two basic frameworks for QC for the theoretical approaches.

One is the periodic orbit theory (POT) pioneered by Gutzwiller, 
which is a semi-classical quantization scheme, applicable
not only to 
the integrable 
but also non-integrable cases 
\cite{gutzwiller71, miller75, gutzwiller80,
gutzwiller82,yellowbook,wintgen88b}. %
The other, random matrix theories (RMT), concerns with the spectral fluctuations 
of QC systems. The level statistics of any QC system is described universally---
independent from the details of the Hamiltonian and only dependent on the symmetry
of it---by one of the three random matrix models, Gaussian 
orthogonal, unitary, and symplectic ensembles 
\cite{wigner51,dyson62a,dyson62b} (Dyson's threefold way) if the statistics is stationary under the shift
of energy, and by additional ensembles \cite{altland97} 
if it is not stationary (Zirnbauer's tenfold way). 
The fact that the RMT give the universal description of QC systems 
was conjectured in \cite{bgs84} (BGS conjecture) and
proved recently after the recognition of a deep link 
between POT and RMT.
That is, the semi-classical POT calculation of the 
level correlations is dominated by the 
contribution of certain PO pairs with like actions \cite{sieber01,sieber02},
and  it was shown that they agree with the sigma model 
calculation mapped from the RMT. 
This beautifully solves the conjecture \cite{heusler07}. 

\subsection{Gutzwiller's trace formula and a reverse use of it}
Since our investigation of the quantum-classical correspondence in AKP is based on the POT,
let us briefly recapitulate the derivation of its core equation, the Gutzwiller's trace formula, for the case of 
a QC system.
It is derived from Feynman's propagator by applying twice the stationary phase approximation.
Firstly from the propagator
\begin{equation}
K(q'',T;q',0)  
=
\braket{q''| 
\exp\left(  - i \frac{\hat{H}}{\hbar }T  \right) 
|q'}
=\int_{q(0)=q'}^{q(T)=q''}{\mathcal D}[q] 
~e^{   \frac{i}{\hbar} \int_{0}^{T}L(q,\dot q) dt},
\label{eq:prop}
\end{equation}
the Green function is defined by  Fourier-Laplace transformation
\begin{equation}
G(q'',q',E)
\equiv 
-\frac{i}{\hbar}\int_{0}^{\infty}e^{i\frac{ET}{\hbar}}
K(q'',T;q',0)dT
\label{eq:green_fn_def}
\end{equation}
where $E$ has infinitesimally small positive imaginary part for convergence.
Using the stationary phase approximation on the path integral side of 
\eqref{eq:prop}, we obtain
\begin{align}
\label{eq:green_fn}
G(q'',q',E)=
\braket{q''| 
\frac{1}{E +i\varepsilon - \hat {H}}| q'} 
\simeq
\sum_{\Gamma_\text{cl}} 
A_{\Gamma} \exp\left(\frac{i}{\hbar}S_\Gamma(q'',q',E) -\frac{i \pi\nu_\Gamma}{2} \right).
\vspace{-1mm}
\end{align}
Now the path integral is
reduced to a sum over classical trajectories $\Gamma_{\rm cl}$s.
The principal function in
\eqref{eq:prop} is replaced by the action $S_\Gamma = {\int_{q'}^{q''}\!\!}_\Gamma
pdq$,
and 
$A_\Gamma$ and $\nu_\Gamma$ are respectively 
Van Vleck determinant and the number of conjugate points of $\Gamma_\text{cl}$. 
Secondly we calculate the response function as a  path-cum-trace of the Green function:
\begin{align}
g(E)
&\equiv\int dq' G(q',q',E).
\label{eq:resp_fn}
\vspace{-1mm}
\end{align}
The $q'$ integration of the first expression in \eqref{eq:green_fn} -the trace of the resolvent operator- gives 
exactly the sum of $1/(E-E_n+i\varepsilon)$ over the energy levels. On the other hand 
the trace of the second may be calculated approximately by the second stationary phase approximation. 
It requires 
\begin{eqnarray}
\left.\left(\frac{\partial S_\Gamma (q'',q',E)}{\partial q''}
\delta q'' 
+
\frac{\partial S_\Gamma (q'',q',E)}{\partial q'}
\delta q'\right)\right|_{\stackrel{q''=q'}{\delta q''=\delta q'}}
=(p''-p')\delta q'
=0. 
\label{eq:stationary_phase_condition}
\vspace{-1mm}
\end{eqnarray}
Thus, the sum over $\Gamma_{\rm cl}$s in \eqref{eq:green_fn} is now reduced to only over all
{\it periodic orbits} in the phase space. 
This sum may be organized as a double summation--
the first sum over all distinct primitive POs and the second
over the repetitions of a given primitive PO, and for each term in the double sum, 
we have to integrate over the initial position $q'$ of the PO. 

To do the $q'$ integral, let us pick an arbitrary point 
$\bar{q}$ on the PO and perform firstly the integration transverse to the PO. 
For the point $q'$ in the transverse neighborhood of  $\bar{q}$
($q'=\bar{q}+\delta q'_\perp$), the action can be approximated as 
\begin{equation}
S_\Gamma(q',q',E)
\approx S_\Gamma(\bar{q},\bar{q},E)
+\frac{1}{2}
\left. \left(
\frac{\partial^2 S_\Gamma}{\partial q''^2_{\perp}}
+2\frac{\partial^2 S_\Gamma}{\partial q''_\perp \partial q'_\perp}
+\frac{\partial^2 S_\Gamma}{\partial q'^2_\perp}
\right) \right|_{\bar{q}} (\delta q'_\perp)^2,
\label{eq:del2_S}
\vspace{-1mm}
\end{equation}
where the term linear in $\delta q'_\perp$ 
has vanished thanks to \eqref{eq:stationary_phase_condition}.
The Gaussian integral over $\delta q'_\perp$ 
yields a determinant factor and the ratio of it to $A_\Gamma$  in \eqref{eq:green_fn} merges into 
the determinant of the monodromy matrix of the periodic orbit. 
It is solely determined by the Lyapunov exponent of the PO 
and independent from the choice of $\bar{q}$.
The zeroth order term in \eqref{eq:del2_S} is also 
(by definition) independent from $\bar{q}$;
$S_\Gamma(\bar{q},\bar{q},E)\equiv S_\Gamma(E)$.
Therefore, the longitudinal integral merely produces the period
of the orbit as a multiplicative factor, and we have reached 
the Gutzwiller's trace formula.  
For a QC system in two dimensions, the trace formula is
\begin{equation}
\vspace{-1mm}
\label{eq:trace_formula}
g(E) =
\sum_n \frac{1}{E-E_n+i\epsilon}
\approx
- \frac{i}{\hbar}
\sum\limits_{\Gamma_{\rm PO}} 
\frac{T_{\Gamma_{\rm PO}}(E)}
{2 \sinh \left( \lambda_{\Gamma_{\rm PO}}/2  \right)}
e^{i S_{\Gamma_{\rm PO}}(E)/\hbar - i \pi \nu_{\Gamma_{\rm PO}}/2}.
\vspace{-1mm}
\end{equation}
Here the sum runs over all periodic orbits (primitive ones and their repetitions) 
and $T_{\Gamma_{\rm PO}}$, 
$S_{\Gamma_{\rm PO}}$, $\nu_{\Gamma_{\rm PO}}$ and 
$\lambda_{\Gamma_{\rm PO}}$, 
are respectively the period, action, number of turning points and Lyapunov exponent 
of each periodic orbit $\Gamma_{\rm PO}$. 
For AKP with azimuthal quantum number $m=0$, this formula is sufficient
to work with. See the discussion at the beginning of Sec~\ref{sec:formulation}.
The density of states (d.o.s.) is directly read off from  
the response function by 
$\rho (E) \equiv 
\sum_n \delta(E-E_n)
=- \frac{1}{\pi } \Im(g(E))$
\footnote{Explicitly writing out the double summation for the POs,
\begin{equation}
 \rho (E)
\approx
\overline{\rho(E)}
+ 
\sum\limits_{\Gamma_{\rm PPO}} 
\sum\limits_{n=1}^\infty
\frac
{ n T_{\Gamma_{\rm PPO}}(E)}
{2 \pi \hbar \sinh\left( n \lambda_{\Gamma_{\rm PPO}} /2 \right)}
\cos n\left( 
S_{\Gamma_{\rm PPO}}(E)/\hbar -  \pi \nu_{\Gamma_{\rm PPO}}/2
\right),
\label{eq:dos}
\end{equation}
where the first term denotes the contribution of zero length orbit.
}.

Let us call the first expression of the trace formula \eqref{eq:trace_formula}
the quantum mechanical (QM) side and the second expression,
the semi-classical approximation for it, the periodic orbit theory (POT) side.  
The QM side is a sequence of sharp 
$\delta-$peaks with a characteristic spectral fluctuation in a certain universality class.
In the POT side, on the other hand, the contribution of a given primitive periodic orbit 
(i.e. the sum over its  repetitions) to $g(E)$ turns out to be a sum of
Lorentzian peaks centered at $E_n$s determined by the relation 
\begin{align}
\label{po_levels}
S(E_n) =2\pi \hbar (n+\nu_{\Gamma_{\rm PPO}}/4)
\end{align}
with widths proportional to the Lyapunov exponent of the PO.   
Therefore, only when the contributions of all primitive POs are summed,
the interference between them will produce the sharp $\delta-$peaks in the QM side
within the semi-classical approximation 
\cite{wintgen88b} 
\footnote{For the integrable systems, the situation is completely different.
A given primitive stable orbit predicts equally spaced {\it sharp}
levels at 
$E_{nk}$ given by 
$S(E_{nk})=2\pi\hbar \left(n+(k+1/2)\beta
+\nu_{\Gamma_{\rm PPO}}/4\right),
$
and there is no interference between such contributions. 
Note the second term with $\beta$ (stability angle divided by $2\pi$)  
is a modification to the Gutzwiller's formula \cite{gutzwiller71}, 
which accounts for the harmonic perturbations
about the periodic orbit \cite{miller75,wintgen88b}.
}.
 
An important test of \eqref{eq:trace_formula} is whether the POT side produces correctly 
the quantum energy levels, when sufficiently large number of periodic orbits
are accounted for. 
This direction (from POT to QM) has been extensively pursued
by Gutzwiller with AKP as a fruitful testing ground \cite{gutzwiller71,gutzwiller77,gutzwiller80,gutzwiller81,gutzwiller82,yellowbook}.

It is possible to organize the POT test in the reverse direction (from QM levels to POT).
Soon after the Gutzwiller's trace formula, a mathematical frame work in this direction was 
presented by Chazarain \cite{chazarain74}, which is an extension of the Poisson's
sum formula, and later on this direction is called {\it inverse quantum chaology} 
and {\it quantum recurrence spectroscopy}\footnote{We have learned these terms
from \cite{berry-speech}.}. One remarkable test in this direction was given by Wintgen 
\cite{wintgen87} in the diamagnetic Kepler problem (DKP)-- 
a hydrogen atom in a uniform magnetic field.  He showed that the long range correlation
in the DKP energy spectrum produces through the trace formula in the reverse direction
peaks due to POs at the right positions on an appropriately scaled action axis.
An important step taken in \cite{wintgen87} is the symmetrization technique given 
by Gutzwiller \cite{gutzwiller82} with which the POT side can accommodate 
the correct symmetry properties of the QM sector with which the QM side is calculated. 
It is the inverse quantum chaology in AKP, which we aim at in this paper. 

\subsection{Anisotropic Kepler problem}
The AKP is a system of an electron, with an anisotropic mass tensor,
bound in a spherically symmetric Coulomb field, which is  
realized by an electron moving around a donor impurity of a semiconductor.
The Hamiltonian (in dimensionless variables) is \cite{gutzwiller82}
\begin{equation}
H=\frac{1}{2\mu} u^2 + \frac{1}{2\nu}( v^2 + w^2) 
- \frac{1}{\sqrt{x^2+y^2+z^2}}~~~~
(\mu \ge \nu,~\mu\nu=1),
\label{eq:hamiltonian}
\end{equation}
where the $x$-axis is the {\it heavy} axis and there remains rotational symmetry 
around the heavy axis due to the crystal symmetry.
 
For the limit $\gamma \equiv \nu/\mu=1$, \eqref{eq:hamiltonian} is the hydrogen
atom (the Kepler problem), one of  the most well-studied  integrable quantum models along 
with the harmonic oscillators. 
With the decrease of the mass anisotropy parameter $\gamma$, 
the classical system increases randomness.
Because of  the Coulombic potential, AKP is not a KAM system \cite{wintgen88a,verbaarschot03}.
with the decrease of $\gamma$ from 1, the phase space does {\it not}
become the admixture of integrable regions (tori)
and ergodic regions, but rather it passes through (around $\gamma \approx 0.75 -0.85$)
an intermediate regime where, due to the abrupt collapse of KAM tori, 
the phase space is filled by Cantori \cite{wintgen88a,verbaarschot03}
(stochastic web of chaos, see e.g. \cite{zaslavsky91}), 
and finally reaches the completely ergodic regime ($\gamma \lesssim 0.3$).
There, the quantum level statistics at the ergodic limit is GOE, because, for AKP, the time-reversal symmetry is respected 
with $T^2=1$ (rather than with $T^2=-1$) and rotational symmetry is also respected. 
In the classical mechanics side, all the periodic orbits are isolated and unstable
\footnote{These estimates of anisotropy ranges come from the level statistics analysis
\cite{wintgen88a, verbaarschot03, ourpaper12,ourpaper13}.
}.

There is a particularly nice feature of AKP. Consider any one of two dimensional AKP orbits 
and code it by a binary code $(\cdots, a_{-1},a_{0},a_{1},\cdots,a_{i},\cdots )$
where $a_i=1~(0)$ if the sign of the $x$ at the $i$-th intersection of the orbit with $x$-axis
is positive (negative). Then, it is proved that there exists,  
for $\gamma <8/9$ (or equivalently $\mu=1/\nu >\sqrt{9/8}$), 
one orbit for one given binary code \cite{gutzwiller77, devaney78}, 
namely AKP at high anisotropy is endowed with such much variety.
Then, a periodic orbit is described by $\{a_i\}$, $(1 \le i \le 2N)$ with $N=1,2,\cdots$. 
Let us call $2N$ the length of the periodic orbit. It is conjectured that the correspondence 
between the code of a periodic orbit and the initial value is  
one to one  
\cite{gutzwiller81,gutzwiller82}
\footnote{\label{footnote:one-to-one-conjecture}
This conjecture is drawn based on shooting for POs with lengths $2N \le 10$
under careful consideration of the orbit symmetries.
We are now able to find POs with $2N \le 20$ and in the way of making exhaustive 
list of them. We have found a plausible explanation of
the conjecture based on the devil's staircase over the two dimensional plane of initial values ($X_0,U_0$) 
with a height function $b(X_0,U_0)$ calculated from the corresponding binary codes
to ($X_0,U_0$) \cite{ourpaper13a}. 
For the use of ($X_0,U_0$), see Gutzwiller \cite{gutzwiller77}. 
}.
Thus, AKP is a best testing ground for the quantum-classical correspondence -- by varying one parameter $\gamma$,
one can observe how it varies, while keeping track of the important ingredients in the classical side, the periodic orbits, by binary codes.

In his seminal paper in 1971\cite{gutzwiller71}, Gutzwiller presented the trace formula
\eqref{eq:trace_formula} for the QC system and applied it to AKP.  
The fundamental periodic orbit (FPO; the distorted Kepler ellipse by mass anisotropy)
was found by solving Hill's equation iteratively and 
the prediction of energy levels using \eqref{po_levels} was shown in rough agreement 
with then available Faulkner's results by perturbative calculation
(the first six levels) \cite{faulkner69} for Si and Ge
\footnote{For silicon 
$\gamma = 0.208$
and for germanium 
$\gamma = 0.0513$.
}.    
Then the endeavour of summing up contributions of all of the periodic 
orbits was accomplished in \cite{gutzwiller80,gutzwiller81,gutzwiller82}.
Here an amazing formula
was utilized, which gives a very good approximation
for the action of a periodic orbit as a function of its binary code.  
Besides a method was given that calculates the energy 
levels of particular quantum subsectors in terms of POT
by the inclusion of periodic orbits which close under
the relevant symmetry transformations \cite{gutzwiller82}.
The real part of the pole positions of the response functions 
turned out in better agreement with the energy levels \cite{faulkner69}
and, most importantly, the imaginary part turned out very small 
(particularly for the parity even sector)
so that it is proved that the interferences between periodic orbits act indeed to 
produce sharp peaks in \eqref{eq:trace_formula}.
Tanner et al. employed the cycle expansion method combined with the functional
equation and obtained even very high-lying energy levels of AKP 
only from $2N \le 8$ POs \cite{tanner91}.
The usual direction of POT trace formula in AKP has been thus
extensively studied. 

\subsection{Inverse quantum chaology in AKP}
In this note, we revisit the anisotropic Kepler problem (AKP), and show that 
the inverse use of the trace formula works perfectly.
Our method is essentially the same with that of Wintgen \cite{wintgen87}. 
We also take full advantage of the Gutzwiller's method of symmetrizing the POT side, 
and we use a naive but very efficient Fourier transformation from the energy 
to action space which undoes the interference and predicts classical periodic orbits
from quantum energy levels one by one. 
Our main formula is \eqref{eq:FT_full}
(and its practical version \eqref{eq:FT_cut}).
The virtue of AKP is its simplicity.
For one thing, the action of the AKP orbit has a simple scaling property  
$S_{\Gamma_{\rm PO}}(E) =\Phi_{\Gamma_{\rm PO}}/\sqrt{-2E}$. 
This makes the experimental inverse test in AKP quite transparent, 
while in DKP the experimental test is somewhat involved
because one has to simultaneously consider the energy levels at different magnetic field strengths
\cite{wintgen94}.  In DKP, we have to also take into account the specific feature that 
the Lyapunov exponent has distinct dependence on the energy and field strength
depending whether the PO is longitudinal or transverse to the magnetic field, 
while in AKP with $m=0$, the POT is two-dimensional and there is only one Lyapunov exponent
to deal with. Finally, one can be sure in AKP for the choice $\gamma=0.2$ (the classic value where the 
Gutzwiller's heroic endeavour was performed) that one is perfectly in the ergodic regime.

For the QM side, we use our data of quantum energy levels extending up to $2 \times 
10^4$ for both even and odd parity sectors.  These data are calculated by the method 
of Wintgen, Marxer and Briggs \cite{wintgen87a}. 
The calculation is performed after a careful test on the necessary choice of the scaling parameter
involved in the formulation. (See, Sec.\ref{subsec:po-and-levels},
footnote$^{\ref{footnote:harmoinic-bases}}$ thereto, and \cite{ourpaper12,ourpaper13}). 

For the POT side, we are now armed by periodic orbit data up to rank $2 N =12$
obtained by an extensive shooting calculation.  (See Sec.\ref{subsec:po-and-levels},   
Table~\ref{tab:number_of_po}, and footnote \ref{footnote:one-to-one-conjecture}). 

We pay special attention in order to perform the test quantitatively.
We check not only the predicted position of the POs but compare the 
the QM and POT sides of the {\it inverse trace formula} \eqref{eq:FT_cut} 
quantitatively with respect to both real and imaginary
part at various choices of the maximum energy levels $n_{\it max}$ to include.
We show that they agree independent from 
$n_{\it max}$ as they should as far as the semi-classical
approximation is correct. 
As far as we know of, this is the first quantitative inverse quantum chaology in AKP.
Furthermore, we show that we can extract not only the action but also the Lyapunov 
exponent of AKP periodic orbits from quantum energy levels.

The organization of the rest of this paper is as follows.
In Section~\ref{sec:formulation}, we present two important ingredients
to facilitate the inverse quantum chaology in AKP. 
One is the Gutzwiller's technique to adapt the POT
so that it can match the symmetry properties of the quantum sector under investigation.
We give explicit formulas for the POT response functions which have proper symmetries. 
The other is the Fourier transformation for the inverse quantum chaology 
using a convenient variable
$u=1/\sqrt{-2E}$ which takes care of the scaling property of the AKP action fully.  
Our main formula is \eqref{eq:FT_full} (and its practical version \eqref{eq:FT_cut}). 
The degeneracy factor $\sigma$ in \eqref{eq:height}
important to perform the quantitative test is discussed.
In Section~\ref{sec:results}, we show our results of inverse quantum chaology in AKP.
In the first part, we compare both sides of the inverse trace formula.
The systematics of the distribution of AKP orbits on the action space
are explained. The characteristic feature of Gutzwiller's amazing action as a 
function of the AKP orbit binary is also discussed.
In the second, we extract periodic orbits in AKP
from quantum data.
We conclude in Section~\ref{sec:conclusion}.
We summarize the WKB method 
which we used to calculate the quantum energy levels of AKP.
The harminic oscillator base calculation to the anisotoropy term in AKP
is new to our knowledge, and which serves a useful formulation for the 
Husimi function calculation in AKP \cite{ourpaper12,ourpaper13}. 


\section{Formulation of Trace Formula in AKP in the Inverse Direction}
\label{sec:formulation}
\subsection{POT response functions with proper symmetries}

Hamiltonian \eqref{eq:hamiltonian} of AKP is symmetric with respect to the rotation around 
the $x$-axis.  Therefore, the angular momentum $L_x$ around the $x$-axis is conserved.
As pointed out by Gutzwiller, the case $L_x =0$
is most  important, since then,  the hard chaos is realized in the classical mechanics.
When $L_x =0$, the orbit of the electron is restricted in a plane 
which includes the $x$-axis. (We choose $xy-$plane as this plane).  
The corresponding QM theory is under the restriction that the azimuthal quantum number 
$m=0$. In the POT side, it is sufficient to work with two dimensional
trace formula \eqref{eq:trace_formula} with the sum over the two 
dimensional periodic orbits and lift it to three dimensional by correctly choosing 
the $\nu_{\Gamma_{\rm PO}}$, while in the QM side the restricted Hilbert space with 
$m=0$ must be calculated with full three-dimensional formulation. 
Moreover, we can restrict the three dimensional quantum theory by the parity $\pi$.
By considering the quantum subsectors $m^\pi=0^+$ and  
$m^\pi=0^-$ separately, and test the quantum-classical correspondence 
sector by sector, we can enhance the accuracy of the test.

Below, we first explain Gutzwiller's idea how to formulate the two
dimensional POT so that it can represent the quantum subsectors $m^\pi=0^+$ and  
$m^\pi=0^-$. 
The strategy is as follows. 
The three dimensional Hamiltonian \eqref{eq:hamiltonian}
is mirror symmetric under the transformation with respect to 
the $yz$-plane. Let us call this $X$-transformation.
The quantum Green function for the parity even (odd)
sector can be constructed by symmetrizing (antisymmetrizing) the Green function 
under the $X$-transformation with respect to the end point of the paths.
Similarly, \eqref{eq:hamiltonian} is also symmetric 
under the rotation around the $x$-axis.
We call this $Y$-transformation (in the two dimensional view). 
The quantum Green function with $m=0$ can be constructed by symmetrizing the Green function 
under the $Y$-transformation with respect to the end point of the paths.
The procedure of symmetrization (antisymmetrization)  can be realized
in terms of POT by introducing the POs that close under the respective transformations. 
Then we give concrete formulas for the response functions which have correct symmetry properties.

\subsubsection{$X$-transformation}

The $X$-transformation is given by 
\begin{eqnarray}
\nonumber
Xq &\equiv& X(x,y,z)~~~~~ Xp \equiv X(u,v,w)\\
&=&(-x,y,z),~~~~~~~~~ =(-u,v,w).
\label{eq:X_tr}
\end{eqnarray}

Firstly let us consider the QM side. 
Because $X^2=1$,   
the eigenfunctions of the Hamiltonian
\eqref{eq:hamiltonian} 
are classified into even or odd functions under 
$X$-transformation, that is 
\begin{equation}
\phi(Xq)=\phi(q),~\psi(Xq)=-\psi(q).
\label{eq:eigenfunction_parity}
\end{equation}
Accordingly the sum for the Green function in the QM side can be
divided into two distinct sums; 
\begin{equation}
G(q'',q',E)
=\sum_j \frac{\phi_j(q'')\phi_j^*(q')}{E-E_j+i\epsilon}
+\sum_k \frac{\psi_k(q'')\psi_k^*(q')}{E-E_k+i\epsilon}.
\label{eq:green_fn_parity}
\end{equation}
Here $j$ and $k$ label respectively even and odd parity quantum states, 
and, as defined in \eqref{eq:resp_fn}, 
the response function is the path-cum-trace  
\begin{eqnarray}
g(E)=\int dq' G(q',q',E).
\label{eq:g}
\end{eqnarray}

By replacing $q''$ by $Xq''$ in \eqref{eq:green_fn_parity} 
and using \eqref{eq:eigenfunction_parity}, 
we obtain  
\begin{eqnarray}
\label{eq:green_fn_parity_x}
\nonumber
G(Xq'',q',E)
&=&\sum_j \frac{\phi_j(Xq'')\phi_j^*(q')}{E-E_j+i\epsilon}
+\sum_k \frac{\psi_k(Xq'')\psi_k^*(q')}{E-E_k+i\epsilon}\\
&=&\sum_j \frac{\phi_j(q'')\phi_j^*(q')}{E-E_j+i\epsilon}
-\sum_k \frac{\psi_k(q'')\psi_k^*(q')}{E-E_k+i\epsilon},
\end{eqnarray}
and the corresponding response function is 
\begin{eqnarray}
g_X(E)=\int dq' G(Xq',q',E).
\label{eq:gX}
\end{eqnarray}
The Green functions for even and odd parity quantum sectors are 
respectively given by 
\begin{eqnarray}
G_e(q'',q',E)
&\equiv&\sum_j \frac{\phi_j(q'')\phi_j^*(q')}{E-E_j+i\epsilon}
=\frac{1}{2}\left[ G(q'',q',E) + G(Xq'',q',E)\right] 
\\
G_o(q'',q',E)
&\equiv&\sum_k \frac{\psi_k(q'')\psi_k^*(q')}{E-E_k+i\epsilon}
=\frac{1}{2}\left[ G(q'',q',E) - G(Xq'',q',E)\right],
\label{eq:green_fn_parity_separated}
\end{eqnarray}
and the desired response functions in the QM side are finally
\begin{eqnarray}
g_e(E)
&=&\sum_j \frac{1}{E-E_j+i\epsilon}
=\frac{1}{2}\left[g(E)+g_X(E)\right]\\
g_o(E)
&=&\sum_k \frac{1}{E-E_k+i\epsilon}
=\frac{1}{2}\left[g(E)-g_X(E)\right].
\label{eq:resp_fn_parity_separated}
\end{eqnarray}

Now, let us discuss the POT side, which is two dimensional. 
For the AKP periodic orbit, the number of conjugate points  
is  twice the number of the intersections of the orbit with 
the $x$-axis and hence $\nu_{\Gamma_{\rm PO}}$ is equal to $4N$ for 
the periodic orbit with the length $2N$.
Thus \eqref{eq:trace_formula} becomes 
\begin{equation}
g(E)
=\sum_n \frac{1}{E-E_n+i\epsilon}
\approx -i\sum_{\Gamma_{\rm PO}} 
\frac{T_{\Gamma_{\rm PO}}(E)}
{2\sinh (\lambda_{\Gamma_{\rm PO}}/2)}
e^{iS_{\Gamma_{\rm PO}} (E)}
\label{eq:trace_formula_akp}
\end{equation}
in appropriate dimensionless variables
used in the Hamiltonian \eqref{eq:hamiltonian}.

\begin{figure}[!tp]
\begin{center}
  \includegraphics[width=80mm,clip]{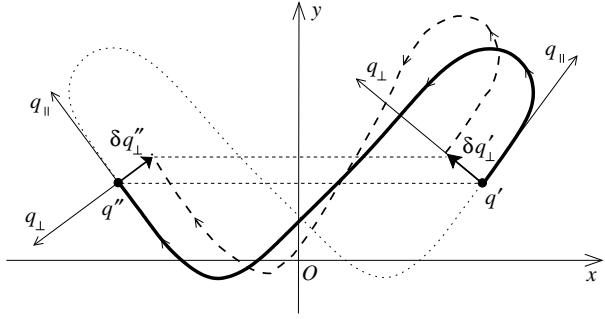}
\end{center}
\caption{
The $X$-periodic orbit (solid curve) 
which closes in $R^2/X$. 
This connects
$q'$ and 
$q''=Xq'$ 
with $p''=Xp'$, 
and its variation (dashed curve) 
with
$\delta q''=X(\delta q'$). 
Note $\delta q''_\perp=-\delta q'_\perp$ 
in the depicted coordinates frames.
}
\label{fig:x_po}
\end{figure}

Now let us consider $g_X(E)$ in \eqref{eq:gX}, which is essential to calculate 
the response functions $g_e(E)$ and $g_o(E)$. 
The condition $q''=Xq'$ implies
 $\delta q''=X \delta q'$, 
and accordingly, the stationary phase condition 
in \eqref{eq:stationary_phase_condition}
becomes
\begin{eqnarray}
\left.\left(\frac{\partial S_\Gamma (q'',q',E)}{\partial q''}
\delta q'' 
+
\frac{\partial S_\Gamma (q'',q',E)}{\partial q'}
\delta q'\right)\right|_{\stackrel{q''=Xq'}{\delta q''=X\delta q'}}
=(Xp''-p')\delta q'
=0.
\label{eq:stationary_phase_condition_sym}
\end{eqnarray}
Thus, not only $q''=Xq'$ but also $p''=Xp'$ hold 
for the orbits which contribute to
$g_X(E)$.
We denote this type of a trajectory as $\Gamma_X$
and call it $X$-periodic orbit.
$\Gamma_X$ is apparently half of a periodic orbit
symmetric with respect to the $y$-axis. See Fig.~\ref{fig:x_po}.
An alternative way of looking at $\Gamma_X$ is that
it is a closed periodic orbit in a orbifold space ${\mathcal R}^2/X$ \cite{robbins89}.
The four characteristic quantities of $\Gamma_X$, namely,
the action, the period, the Lyapunov exponent, and the number of conjugate points, are all half of those of the corresponding full periodic orbit.

Now, for $X$-periodic orbit,  $q'' = X q'$
implies $\delta q''_\perp=-\delta q'_\perp$, and 
instead of \eqref{eq:del2_S}, 
we obtain 
\begin{equation}
S_\Gamma(Xq',q',E)
=S_\Gamma(X\bar{q},\bar{q},E)
+\frac{1}{2}\left(
\frac{\partial^2 S_\Gamma}{\partial q''^2_{\perp}}
-2\frac{\partial^2 S_\Gamma}{\partial q''_\perp \partial q'_\perp}
+\frac{\partial^2 S_\Gamma}{\partial q'^2_\perp}
\right) (\delta q'_\perp)^2.
\label{eq:del2_S_sym}
\end{equation}
The negative sign in the second term of the right hand side then requires 
the replacement of the hyperbolic sine function by the hyperbolic cosine function 
in the amplitude factor in the trace formula 
\eqref{eq:trace_formula}.
Furthermore,  the length 
of the full periodic orbit, symmetric with respect to the y-axis,
is $2N$ with $N$ itself being even in order to guarantee the symmetry
\footnote{If the orbit is symmetric with respect to the both x-axis
and y-axis, $N$ can be any odd integer number.
This exception includes, for instance, the fundamental periodic orbit (FPO).  
}.
Accordingly the number of the conjugate point $\nu_{\Gamma_X}$ 
is $2N$ (half of $4N$) with an even $N$, and the  phase factor,   
depending on $\nu_{\Gamma_X}$ should be omitted.
In total, we obtain an explicit formula for the POT side of $g_X(E)$ as
\begin{equation}
g_X(E)
\approx -i\sum_{\Gamma_X} \frac{T_{\Gamma_X}(E)}
{2\cosh (\lambda_{\Gamma_X}/2)}e^{iS_{\Gamma_X} (E)}.
\label{eq:trace_formula_X}
\end{equation}
This important issue will be tested quantitatively in Sec.~\ref{sec:results}.

\subsubsection{$Y$-transformation}

In order to manufacture the three dimensional rotational 
symmetry around the $x$-axis 
in terms of the  two dimensional trace formula, 
we have to select the orbits $\Gamma_Y$ which are closed 
in the orbifold space ${\mathcal R}^2/Y$.
We call this type of half orbits $Y$-periodic orbit. 

For the $Y$-periodic orbit, 
$\delta q''_\perp=\delta q'_\perp$
and  there does not occur the change of sign 
which occurred in \eqref{eq:del2_S_sym}, 
and there is no need for interchanging the hyperbolic functions  
in the amplitude factor. 
For the length of the full periodic orbit 
whose half is $\Gamma_Y$,  
$N$ is an odd natural number.  
The number of conjugate points $\nu_{\Gamma_Y}$ 
is $2N$ (half of $4N$) with odd  $N$. \\ 
In total we obtain an explicit semi-classical formula for $g_Y(E)$
\begin{equation}
g_Y(E)
\approx i\sum_{\Gamma_Y} \frac{T_{\Gamma_Y}(E)}
{2\sinh (\lambda_{\Gamma_Y}/2)}e^{iS_{\Gamma_Y} (E)}
\label{eq:trace_formula_Y}
\end{equation}
with $i$ in the front.

\subsubsection{Combination of $X$- and $Y$- transformations}

By considering the $X$- and $Y$- transformations 
at the same time, 
we can calculate individually the even and odd 
response functions under $X$- transformation,  
both for $m=0$ subspace as 
\begin{align}
\label{eq:full-resp_fn}
\nonumber
g_e^{m=0}(E)
&=\frac{1}{4}\left[ g(E)+g_Y(E)+g_X(E)+g_{XY}(E)\right]\\
g_o^{m=0}(E)
&=\frac{1}{4}\left[ g(E)+g_Y(E)-g_X(E)-g_{XY}(E)\right].
\end{align}
The new term 
\begin{align}
g_{XY}(E)=\int dq' G(XYq',q',E)
\label{eq:gXY}
\end{align}
in the POT side can be handled in the same manner as above, 
and we obtain
\begin{equation}
g_{XY}(E)
\approx i\sum_{\Gamma_{XY}} \frac{T_{\Gamma_{XY}}(E)}
{2\cosh (\lambda_{\Gamma_{XY}}/2)}e^{iS_{\Gamma_{XY}} (E)}
\label{eq:trace_formula_XY}.
\end{equation}
Here the orbits $\Gamma_{XY}$ are closed 
in the orbifold space ${\mathcal R}^2/(X\cdot Y)$ 
and we call these half orbits $XY$-periodic orbits.
For the length $4N$ of the full periodic orbit 
whose half is $\Gamma_{XY}$, $N$ is an odd number.  
The number of conjugate points $\nu_{\Gamma_{XY}}$ 
is $2N$ (half of $4N$) with odd $N$.

\subsection{Inverse quantum chaology: use of a  variable $u= {1}/{\sqrt{-2E}}$}
\label{use_of_u}

If we follow the trace formula in the direction (from POT to QM),
the density of state (d.o.s.) will be produced (within the semi-classical approximation) 
but the interference between the PO contributions necessarily forbids to investigate the 
contributions of periodic orbits one by one.
Instead in the inverse direction (from QM to POT),
we are going to extract the information of periodic orbits from the d.o.s.
one by one. For this purpose, we focus on the scaling property 
of AKP action.
 
Because of the Hamiltonian \eqref{eq:hamiltonian} is 
homogeneous with respect to coordinates and momenta, 
the action of AKP has a remarkable scaling property 
\begin{equation}
S_{\Gamma_{\rm PO}} (E)=\frac{1}{\sqrt{-2E}}
\Phi_{\Gamma_{\rm PO}}.
\label{eq:akp_action}
\end{equation}
Here $\Phi_{\Gamma_{\rm PO}}$ is the value of the action of 
$\Gamma_{\rm PO}$
at $E=-1/2$ and it is an energy independent constant 
distinguishing the periodic orbits.  Apart from this factor
all the periodic orbits have the same energy dependence 
($\propto 1/\sqrt{-2E}$).
From this observation, we introduce a convenient variable
\begin{equation}
u\equiv\frac{1}{\sqrt{-2E}}. 
\label{eq:u}
\end{equation}

Now we change the variable from $E$ to $u$
in each side of \eqref{eq:trace_formula_akp}.  
In QM side, we have simply 
\begin{equation}
\frac{1}{E-E_n+i\epsilon}
=\frac{du}{dE}\frac{1}{u-u_n+i\epsilon}.
\label{eq:e_to_u}
\end{equation}
In the POT side, 
the period is given by $T(E)=d S(E)/dE$
and we have 
\begin{equation}
T_{\Gamma_{\rm PO}} (E)
=\frac{d u}{dE}\frac{d}{du}S_{\Gamma_{\rm PO}}
=\frac{d u}{dE}\Phi_{\Gamma_{\rm PO}}.
\label{eq:akp_period}
\end{equation}
Because the factor $du/dE$ in \eqref{eq:e_to_u}and \eqref{eq:akp_period}
cancels out, we obtain a concise formula for the response function in 
terms of  $u$; 
\begin{equation}
g(u)\equiv \sum_n \frac{1}{u-u_n+i\epsilon}
\approx -i\sum_{\Gamma_{\rm PO}} 
\frac{\Phi_{\Gamma_{\rm PO}}}
{2\sinh (\lambda_{\Gamma_{\rm PO}}/2)}
e^{iu\Phi_{\Gamma_{\rm PO}}}.
\label{eq:trace_formula_in_u}
\end{equation}

Let us consider the Fourier-Laplace transformation 
from $u$ to the conjugate variable $\phi$
\begin{equation}
D(\phi) \equiv \int_0^\infty g(u) e^{-iu\phi} du.
\label{eq:fourier}
\end{equation}
Then  we obtain from \eqref{eq:trace_formula_in_u}
\begin{equation}
D_{\rm QM}(\phi) 
\equiv \sum_{n=1}^{\infty} e^{-i u_n \phi}
\approx 
D_{\rm POT}(\phi) \equiv \sum_{\Gamma_{\rm PO}} 
\frac{\Phi_{\Gamma_{\rm PO}}}
{2\sinh (\lambda_{\Gamma_{\rm PO}}/2)}
\delta \left(\phi - \Phi_{\Gamma_{\rm PO}} \right).
\label{eq:FT_full}
\end{equation}

In the POT side, $D_{\rm POT}(\phi)$ has
delta peaks localized at $\phi=\Phi_{\Gamma_{\rm PO}}$. 
Hence there is no interference between contributions 
from  periodic orbits. It is, so to speak, {\it the (weighted) density of the periodic orbits}.
On the other hand, in $D_{\rm QM}(\phi)$, 
the pseudo energy levels $e^{-i u_n \phi}$
distributed on the unit circle on the complex plane
with fluctuations interfere each other. 
Eq. \eqref{eq:FT_full} predicts that $D_{\rm QM}(\phi)$ should have peaks at 
$\phi=\Phi_{\Gamma_{\rm PO}}$ 
by this interference. 
While in \eqref{eq:trace_formula} it is the POT side 
that produces peaks by the interference between periodic orbits,
in \eqref{eq:FT_full} it is the QM side that produces peaks
by the interference between pseudo energy levels.
The role of left and right hand side is interchanged between two trace formulas
\eqref{eq:trace_formula} and \eqref{eq:FT_full}. 

Practically the available energy levels are 
 limited to $n_{\rm max}$, and we have to 
introduce a corresponding cut off  
$u_{\rm max}\equiv u_{n_{\rm max}}$.
Thus we consider
\begin{equation}
D^{\rm cut}(\phi) = \int_0^\infty \theta(u_{\rm max}-u) g(u) e^{-iu\phi} du.
\label{eq:fourier-cut}
\end{equation}
We obtain then
\begin{equation}
D^{\rm cut}_{\rm QM}(\phi)
\equiv \sum_{n=1}^{n_{\rm max}} e^{-i u_n \phi}
\approx 
D^{\rm cut}_{\rm POT}(\phi)
\equiv
\sum_{\Gamma_{\rm PO}} 
\frac{\Phi_{\Gamma_{\rm PO}}}
{2\sinh (\lambda_{\Gamma_{\rm PO}}/2)}
\tilde{\delta} \left(\Phi_{\Gamma_{\rm PO}}-\phi \right),
\label{eq:FT_cut}
\end{equation}
with a smeared delta function
\begin{equation}
	\tilde{\delta}(w)
	=\frac{1}{2\pi }\frac{e^{iw u_{\rm max}}-1}{i w}
	\label{eq:delta-like}
\end{equation}
with a peak at $w=0$ in the real part. 
The formula \eqref{eq:delta-like} provide us with a new test
by comparing $D^{\rm cut}_{\rm QM}(\phi)$ and  
$D^{\rm cut}_{\rm POT}(\phi)$
with freely chosen  $n_{\rm max}$.

If we choose sufficiently large  $n_{\rm max}$, 
the interference between peaks in \eqref{eq:FT_cut}
is strongly suppressed.  
In this situation, the POT side predicts that
the peak in  
$D^{\rm cut}_{\rm QM}(\phi)$ 
should appear at $\phi=\Phi_{\Gamma_{\rm PO}}$ for every periodic orbit, and the predicted ($u_{\rm max}$-dependent) height of the 
peak of the real part is 
\begin{equation}
	\frac{\sigma\Phi_{\Gamma_{\rm PO}}}
	{2\sinh(\lambda_{\Gamma_{\rm PO}}/2)}
	\frac{u_{\rm max}}{2}.
	\label{eq:height}
\end{equation}
 Here $\sigma$ is the number of the periodic orbits
which are degenerate at the same $\Phi_{\Gamma_{\rm PO}}$ 
due to the symmetries of the Hamiltonian.
This degeneracy factor $\sigma$ is listed in Table I 
in \cite{gutzwiller81} and we reproduce it 
at the upper row in our Table~\ref{tab:number_of_po}.
The lower row of Table~\ref{tab:number_of_po} lists the number of periodic orbits 
with a given symmetry in the case of orbit length $2N=12$ as an example.
The sequence of the above calculation is 
\begin{align*}
g(E) \;\text{in}\;\eqref{eq:trace_formula_akp} 
\rightarrow \;g(u)\;\text{in}\;\eqref{eq:trace_formula_in_u} 
\xrightarrow{\rm\;Fourier\;Tr.\;in\;\eqref{eq:fourier}\;} D_{\rm POT}(\phi)\;\text{in}\;
{\rm the\;inverse\;trace\;formula}\eqref{eq:FT_full},
\end{align*}
and
\begin{align*}
g(E) \; 
\rightarrow \;g(u)\;\xrightarrow{\rm\;Fourier\;Tr.\;with\;cut\;in\;\eqref{eq:fourier-cut}\;} 
D_{\rm POT}^{\rm cut}(\phi)\;\text{in}\;{\rm the\;inverse\;trace\;formula\;under\;cut}\eqref{eq:FT_cut}.
\end{align*}
For the inverse quantum chaology
from the quantum subsector $m^\pi=0^+$ and $m^\pi=0^-$,
we need  also $D_X$, $D_Y$, and $D_{XY}$ in POT. (See equations \eqref{eq:D_fn}
and \eqref{eq:full-resp_fn}).
These  can be calculated just in the same scheme 
starting from $g_X(E)$, $g_Y(E)$ and $g_{XY}(E)$ 
in \eqref{eq:trace_formula_X},
\eqref{eq:trace_formula_Y} 
and \eqref{eq:trace_formula_XY}, respectively.

\begin{table}[b]
\begin{center}
\caption{Types of periodic orbits and corresponding 
degeneracy factor $\sigma$ (\cite{gutzwiller81}) (upper row).
Also the number of periodic orbits in each symmetry type 
is listed for length $2N=12$ (lower).
For instance, there are 60 non-self-retracing periodic orbits that are neither 
$X$- nor $Y$-symmetric
(each has $\sigma=2^3$).  
}
\begin{tabular}{ccccp{20pt}cccc}
\hline
\multicolumn{4}{c}{Not self-retracing}&&
\multicolumn{4}{c}{self-retracing}\\[-1mm]
\multicolumn{1}{c}{(i) }
&\multicolumn{1}{c}{(ii)}
&\multicolumn{1}{c}{(iii)}
&\multicolumn{1}{c}{(iv)}
&&
\multicolumn{1}{c}{(i) }
&\multicolumn{1}{c}{(ii)}
&\multicolumn{1}{c}{(iii)}
&\multicolumn{1}{c}{(iv)}\\
\multicolumn{1}{c}{No sym. }
&\multicolumn{1}{c}{X-sym.}
&\multicolumn{1}{c}{Y-sym.}
&\multicolumn{1}{c}{X,Y-sym.}
&&
\multicolumn{1}{c}{No sym.}
&\multicolumn{1}{c}{X-sym.}
&\multicolumn{1}{c}{Y-sym.}
&\multicolumn{1}{c}{X,Y-sym.} \\
\hline
\multicolumn{1}{c}{8 }
&\multicolumn{1}{c}{4}
&\multicolumn{1}{c}{4}
&\multicolumn{1}{c}{2}
&&
\multicolumn{1}{c}{4}
&\multicolumn{1}{c}{2}
&\multicolumn{1}{c}{2}
&\multicolumn{1}{c}{1}\\
\hline\\[-4mm]
60& 15& 28& 2&& 12& 4& 1& 0\\[1mm]
\hline
\end{tabular}
\label{tab:number_of_po}
\end{center}
\end{table}

As the degeneracy factor $\sigma$ plays a crucial role in the quantitative test of the 
inverse trace formula \eqref{eq:FT_full} (\eqref{eq:FT_cut})
let us briefly discuss it before concluding this subsection.
We need to consider three symmetry transformations (acting on the periodic orbits)
in two dimensions;
$X$-transformation, $Y$-transformation, and time reversal ${\mathcal T}$.

Firstly, let us consider the $X$- and $Y$-transformations.
Either of them can produce from a given periodic 
orbit a partner periodic orbit with the same period 
and a respectable solution of the Hamilton equation of motion.
Now, the key point is whether the partner 
is distinct or not. This depends on the symmetry of the shape of 
the periodic orbit.
 There are four types of periodic orbits,
namely, 
(i) those which have neither $X$- nor $Y$-symmetry, 
(ii,iii) those which have either $X$- or $Y$-symmetry but not both, 
and (iv) those which have both symmetries.
If a periodic orbit is type (i), 
it is associated with $2 \times 2 =4$
distinct partners of the same period and we must  give a factor four 
multiplicatively in $\sigma$. 
Similarly, for type (ii,iii), and (iv)
we must give two and one in $\sigma$ respectively.

As for the time reversal ${\mathcal T}$, we should note that
there are {\it self-retracing} periodic orbits, which retrace  themselves
after passing the turning point. (See, for instance, the 
orbit in the third inset of Fig. \ref{fig:ft3d}). 
The time reversal transformation produces from a given periodic orbit 
its partner rotating in the reverse
direction and for this pairing we should, except for the 
self-retracing type, 
count a factor of two multiplicatively  
in $\sigma$. If the orbit is self-retracing, 
the pairing is immaterial since it only amounts to another choice of the starting point and the multiplicative factor is one .   

As a concrete example, let us consider the FPO, 
which is a variation of the Kepler ellipse 
under the mass anisotropy. It is both $X$- and $Y$-symmetric
and hence it is a singlet with respect to these symmetry. 
As for the time-reversal, it is not self-retracing
and hence we must count two for time-reversal.
Therefore, $\sigma=1 \times 2$ for the FPO.

As another concrete example, let us consider 
the maximally-degenerate case, $\sigma_{\rm max}=2^3=8$.
This is the case when the orbit has neither $X$-, $Y$-symmetry nor 
the self-retracing property.

\section{Results}
\label{sec:results}
\subsection{The set up of the inverse quantum chaology in AKP}
\label{subsec:po-and-levels}

We choose for the anisotropy $\mu=\sqrt{5}$ 
(or equivalently $\gamma=0.2$ in the notation
of  \cite{wintgen87a, wintgen88a}).
This anisotropy is realized by the electron in
silicon semi-conductor.
At this anisotropy, the electron classical orbit is perfectly chaotic
and the unstable periodic orbits are isolated each other in the classical
phase space. The classical theory of AKP at $\mu=\sqrt{5}$ was 
extensively examined by Gutzwiller
and both action and Lyapunov exponent 
of periodic orbits with length $2N$ up to ten 
are  tabulated in \cite{gutzwiller81}.
At $2N=2,4,6,8,10$ there are respectively $2, 4, 8, 18,44 $
 independent periodic orbits, each has its own $\sigma$ according to its symmetries as discussed
in subsection \ref{use_of_u}.
In order to establish our test of quantum-classical correspondence, 
we have extended 
\footnote{
The procedure is essentially shooting but with a special attention
to the appearance of flat directions in the 
chi-squared function expressing the amount of misfit in the shooting.  
This flat direction corresponds to the section of the 
surface of the devil's stair case over the $(X_0,U_0)$ at the height 
of the binary code of the periodic orbit.  See also footnote \ref{footnote:one-to-one-conjecture}
and \cite{ourpaper13a}
.}
the Gutzwiller table 
to include altogether 122 independent periodic orbits 
at $2N=12$.
This is sufficient to support the analysis for the region 
$\phi=0-15$. 
(See the discussion in subsection  
\ref{subsect:investigation_of_QCC} 
below and Fig. \ref{fig:ft3d}).
The numbers of periodic orbits subject to the symmetry type (i), (ii), (iii) and (iv) 
are tabulated in Table~\ref{tab:number_of_po} along with
the degeneracy factor $\sigma$ for each type of PO.

As for the energy levels in the QM side, 
we have worked out up to around $10^4$ energy levels from the ground state 
at $\mu=\sqrt{5}$ ($\gamma=0.2$) using the matrix diagonalization method 
with Sturmian basis developed 
by Wintgen, Marxer, and Briggs \cite{wintgen87a}.
This is not a perturbation calculation but rather an exact calculation
in the sense that the approximation only comes from the practical 
limitation of the size $N$ of the (suitably deformed) Hamiltonian matrix.
We derived in our earlier work  two useful formulas \cite{ourpaper12, ourpaper13}; 
the most optimized value of the scaling parameter $\varepsilon$ involved in the 
WMB formulation is $\varepsilon \approx -\gamma/4$ 
and,  with this $\varepsilon$,
\begin{align}
\label{Reff}
{\mathcal R}_{\rm eff}
\equiv 
\frac{\text{number  of reliable energy levels}}{\text{matrix size N}}\approx \sqrt{\gamma}
=\frac{1}{\mu}. 
\end{align}
Eq. \eqref{Reff} implies that the calculation of higher $\mu$ (lower $\gamma$) becomes
increasingly harder. 
For instance, a large size calculation of $10^4$ energy levels, the same number of levels with us, 
was previously performed at $\gamma=0.8$ by WMB. 
But, according to \eqref{Reff}, ${\mathcal R}_{\rm eff}=0.9$ and $0.46$
at $\gamma=0.8$ and $0.2$ respectively. Thus we had to use 
the Hamiltonian matrix with $N$ being roughly twice of that of WMB and  
the necessary CPU memory is about four times
\footnote{
\label{footnote:harmoinic-bases}
We have also formulated WMB diagonalization in terms of 
tensored two-dimensional harmonic-oscillator function basis in order to study
quantum-classical correspondence in the phase space, because 
the harmonic oscillator basis makes Husimi function calculation 
much easier. The comparison between the calculations in two independent basis
has in turn given us  guarantee on the accuracy of both calculations.
In the new basis, 
$\hat{\varepsilon}=\gamma$ and 
$\hat{R}=\sqrt{\gamma}/2$ 
\cite{ourpaper12, ourpaper13}. 
}.
\\

\subsection{
Investigation of the quantum-classical correspondence
by the comparison between $D_{\rm QM}(\phi)$ and $D_{\rm POT}(\phi)$
}\label{subsect:investigation_of_QCC}

\begin{figure}[!bp]
\begin{center}
  \includegraphics[width=160mm,clip]{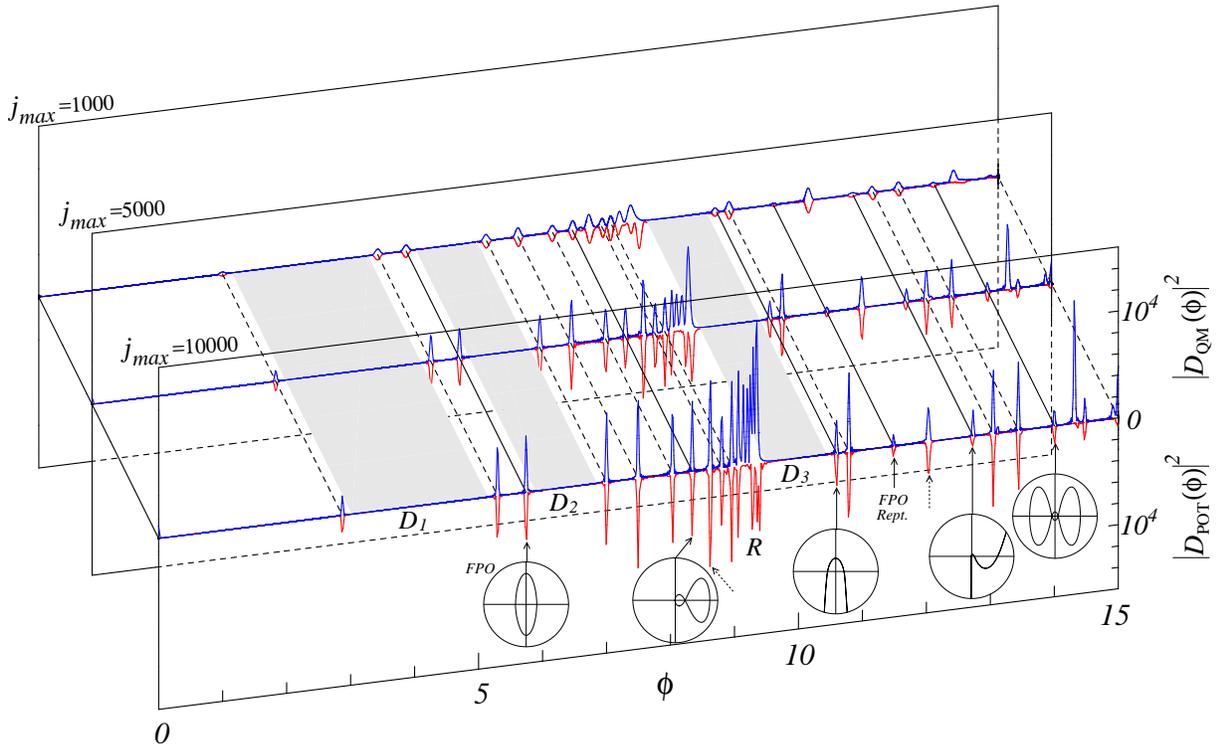}
\end{center}
\caption{
Comparison of QM and POT side of
\eqref{eq:FT_cut} at anisotropy $\mu=\sqrt{5}$ and for 
the quantum sector $m=0$, parity even.
In each of three vertical panels with distinct 
$j_{\rm max}$,
$|D_{\rm QM}(\phi)|^2$ (upper)
and $|D_{\rm POT}(\phi)|^2$ (lower)
are compared after subtracting Thomas-Fermi 
contribution to avoid extraneous peak around 
$\phi=0$. 
Clustering of peaks (${\mathcal R}$) and deserts of peaks (${\mathcal D}_{1,2,3}$)
 are observed.
The POT side is calculated from 2-dim. orbits
($2N \le 12$); six full POs (including repetition of FPO) 
and ten half POs. 
Peaks due to the former (the latter) are connected 
by solid (dashed) lines perpendicular to panels.
Insets show full POs 
in the order of increasing action
$\Phi_{\Gamma_{\rm PO}}$, 
11(FPO), 22, 23, 33, 35 respectively \cite{gutzwiller81}.
There are two peaks (indicated by dashed arrows without insets),
each is produced by two periodic orbits
with almost degenerated actions.
}
\label{fig:ft3d}
\end{figure}

  In Fig. \ref{fig:ft3d} we compare $|D_{\rm QM}(\phi)|^2$ and 
$|D_{\rm POT}(\phi)|^2$ (the latter is flipped vertically for the sake of comparison) 
for the quantum subsector $m^\pi=0^+$.
(Hereafter we drop the superscript `cut' for simplicity).
The three panels show the results at $j_{\rm max}=10^3, 5 \times 10^3, 10^4$ respectively. 
We find that, with the increase of $j_{\rm max}$, the peaks become higher
and the widths become narrower. This trend is common to the QM and POT sides
and at every $j_{\rm max}$ the agreement between
$|D_{\rm QM}(\phi)|^2$ and 
$|D_{\rm POT}(\phi)|^2$ is remarkable.  This is just as it should be
(see the subsection \ref{use_of_u}).

Note that the graphs of $|D_{\rm QM}(\phi)|^2$ are the
result of the sum of the pseudo-energy terms $\exp{(i u_j \phi)}$
as given in the QM side. (See \eqref{eq:FT_cut}).
Here, it is checked that the energy levels $E_j$ (recall $u_j \equiv 1/\sqrt{-2E_j}$) 
are subject to GOE at this high anisotropy. The statistics,
for instance, $\Sigma_2$ and $\Delta_3$, are well described by the GOE \cite{wintgen88a}.
Even knowing that the Gutzwiller's trace formula should give a correct 
semi-classical description, it is still amazing that 
the sum of the random $\exp{(i u_j \phi)}$s
on the unit circle on the complex plane produces 
by the interference the peaks just at the very locations in $\phi$
with the heights and widths just prescribed by the POT side. 
(For instance, if we plot the distribution of $\exp{(i u_j \phi)}$s as a histogram on the circle,
we can see no structure in it. Neither, the mock levels created by Thomas-Fermi formula
shows no peaks at all on the $\phi$-axis. The fluctuation of the levels 
creates the peaks at the right positions with the correct widths as expected by the actions and 
Lyapunov exponents of the periodic orbits.) 

The Gutzwiller's trace formula works surprisingly well --- even magically. 
This point may be stated also in a reverse way;
it may be said that the QM levels amazingly\;{\it know} the information of 
periodic orbits.

Now let us examine Fig. \ref{fig:ft3d} in more details.
The distribution of the peaks in the region $\phi=0-15$
shown in  Fig. \ref{fig:ft3d} has interesting three features;\\
(i) Peaks appear usually isolated each other. \\
(ii) But a strong accumulation is observed around $\phi \in (8.5, 9.5)$. \\
(iii) There are three remarkable gaps where no peaks appear. \\
We explain these features  in terms of the POT as follows.

\vspace{2mm}
(i) Each of the isolated peaks can be attributed to a corresponding
periodic orbit with the same period. In the region $\phi = 0-15$ 
depicted in the figure, there are altogether 16 isolated peaks. 
Among them, six peaks are attributed to the full orbits which are  
respectively displayed in the insets. Other 10 peaks are attributed 
to half periodic orbits $\Gamma_X, \Gamma_Y, \Gamma_{XY}$. 

The check in the reverse way is also successful.
In fact all the periodic orbits  up to $2N=10$ correspond to
all the peaks in the QM side and nothing else. 
The $2N=12$ periodic orbits (those we have newly worked out)
correspond to peaks 
with period $\phi$ above the region $\phi=0-15$ and in
the accumulation region $\phi \in (8.5, 9.5)$. See further discussion
concerning the feature (iii) below.
This successful attribution of all of the isolated peaks in the QM side 
to the full and half orbits in the POT side is a direct evidence 
of the trace formula combined with the orbifold symmetrization discussed in subsection \ref{use_of_u}.

\vspace{2mm}
(ii) Now let us discuss the accumulation of peaks in the region
${\mathcal R} \equiv \{\phi \in (8.5,~9.5)\}$ 
produced by the energy level data 
in the QM side.  With the increase  $j_{\rm max}$,
more peaks are resolved in this accumulation.
This accumulation (the clustering of peaks) produced in the QM
side can be succinctly explained from 
the POT side as follows.
In AKP there is a special class of periodic orbits
--- infinitely many periodic orbits,
each evolving along the heavy $x$-axis
with a small-amplitude oscillation in the transverse $y$-direction.
All of the periodic orbits in this class or their
half orbits tend to have nearly the same action
$\Phi_{\Gamma_{\rm PO}}$. 
The upper bound of these $\Phi_{\Gamma_{\rm PO}}$
can be estimated by considering the one dimensional orbit evolving precisely on the 
x-axis connecting the two turning points at the
kinematical boundary ($x=\pm 2$), that is,
the limit of no oscillation in the $y$-direction.
The action can be calculated analytically and 
$\Phi_{\rm 1dim}=2\pi \sqrt{\mu}\simeq 9.396$.
Therefore, the accumulation of peaks produced in ${\mathcal R}$
by the QM energy level data 
is just as predicted from the POT side consideration.
Equivalently we may say that the QM side amazingly
{\it knows} the accumulation of peaks in ${\mathcal R}$.

\vspace{2mm}
(iii)
As for the {\it deserts} 
${\mathcal D_1}$, ${\mathcal D_2}$, and ${\mathcal D_3}$ 
with no peaks at all  in the QM side, we have firstly checked explicitly that, 
consistently, there exit  no periodic orbits with the action in these ranges
in our collections of periodic orbits up to $2N=12$.
Does any one of periodic orbits with higher lengths come with its action 
in these ranges?
This is a hard question since it is not tractable 
to exhaustively search for 
periodic orbits and examine their actions. 
But we have two supporting evidences for the non existence of such periodic orbits. For one thing there is a plausible tendency that higher length orbits come 
with larger action (i.e. period) and it seems that $2N=12$ is  already 
the highest length that can give actions in the range $\phi \le 15$.  
A better evidence is given by an amazing formula constructed by
Gutzwiller for AKP, which approximately gives $\Phi_{\Gamma_{\rm PO}}$ 
from the Bernoulli sequence $(a_i, ~i=1,\cdots, 2N)$
of $\Gamma_{\rm PO}$
\cite{gutzwiller80, gutzwiller81, gutzwiller82}.
If this formula, derived from the periodic orbits up to 
$2N=10$, can be also used for higher length orbits
\footnote{At least we have checked that it nicely approximates our
new $2N=12$ data except for the special class of periodic orbits
responsible for the accumulation of peaks in ${\it \mathcal R}$
discussed in the feature (ii) above.
}
, 
we can draw a conclusion that there are no higher length periodic orbits
in the desert regions.
In short, the QM side knows not only the accumulation of peaks in 
${\mathcal R}$
but also the deserts ${\mathcal D_1}$, ${\mathcal D_2}$ and ${\mathcal D_3}$.\\


\subsection{Extraction of the information of periodic orbit}

\begin{figure}[!tp]
\begin{center}
  \includegraphics[width=80mm,clip]{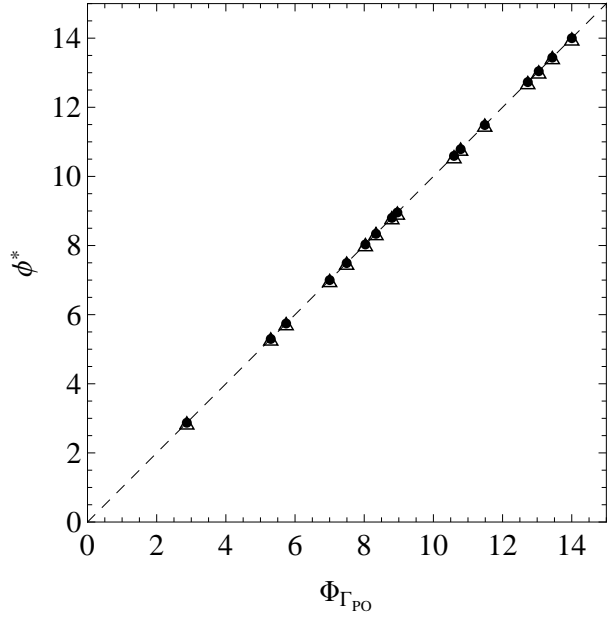}
\end{center}
\caption{
Scatter plot of ($\Phi_{\Gamma_{\rm PO}}$,$\phi^{*}$)
for the sixteen peaks in the range
$\phi \in (0, 15)$,
where $\Phi_{\Gamma_{\rm PO}}$ is the action of each 
PO and $\phi^{*}$ is the corresponding location of the peak
in the QM side of \eqref{eq:FT_cut}.
The QM-POT correspondence is accurate
for both $j_{\rm max}=1000$ (white triangles), 
and $10000$(black dots).
}
\label{fig:s_comparison}
\end{figure}
\begin{figure}[!tbp]
\begin{center}
  \includegraphics[width=80mm,clip]{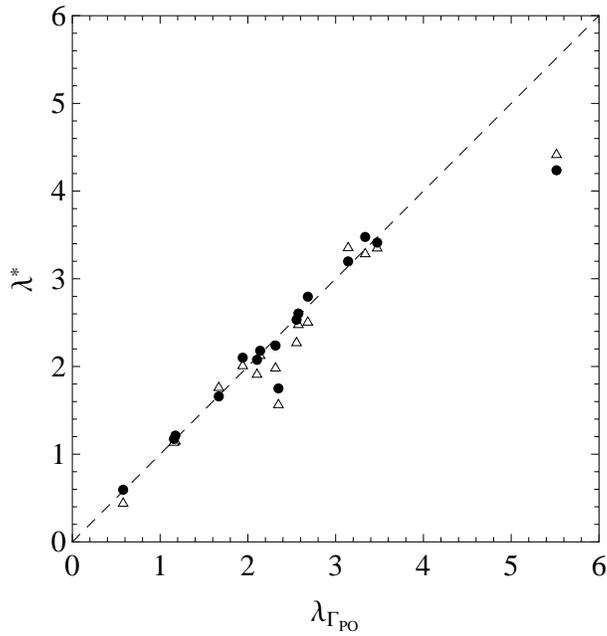}
\end{center}
\caption{Scatter plot of  ($\lambda_{\Gamma_{\rm PO}}$, $\lambda^*$)
for the sixteen POs in the range $\phi\in (0,15)$, where 
$\lambda_{\Gamma_{\rm PO}}$ is the
unstable exponent of a PO,
and $\lambda^*$ is its estimate
from QM side. 
Symbols are the same with 
Fig. \ref{fig:s_comparison}.
Estimate is the better with the larger $j_{\rm max}$. 
}
\label{fig:us-exp_comparison}
\end{figure}

In the previous subsection \ref{subsect:investigation_of_QCC} 
we have found that, in a way, 
the QM side of \eqref{eq:FT_full} (or \eqref{eq:FT_cut}) {\it knows}
the periodic orbits. 
Let us now quantitatively check this issue in 
Fig. \ref{fig:s_comparison} and 
\ref{fig:us-exp_comparison}.

In Fig. \ref{fig:s_comparison}  
the horizontal axis is for $\Phi_{\Gamma_{\rm PO}}$
of the sixteen periodic orbits in the range
$\phi \in (0, 15)$ involved in the POT side of \eqref{eq:FT_cut},
and the vertical axis is for the location of the sixteen peaks 
at $\phi^{*}$ in $|D_{\rm QM}(\phi)|^2$ in the QM side.
Then we combine $\Phi_{\Gamma_{\rm PO}}$ and the corresponding
$\phi^{*}$ into a point  ($\Phi_{\Gamma_{\rm PO}}$,$\phi^{*}$)
using the attribution disscussed in the feature (i) in 
subsection \ref{subsect:investigation_of_QCC} and 
plot the position in the $xy$-plane. 
Note that this attribution--one to one correspondence--
is supported in a triple way, that is, the agreement of the action (peak location),
the height and  the width of the peak. We observe that the
correspondence is extremely accurate; both sets of the sixteen points, 
constructed  from the $j_{\rm max}=1000$ and $10000$ respectively,
equally lie on the line $y=x$.  The correspondence is robust
for the choice of $j_{\rm max}$.

\vspace{2mm}
Encouraged by this success let us attempt to deduce the 
information of the Lyapunov exponents 
 $\lambda_{\Gamma_{\rm PO}}$ for the above sixteen periodic orbits
of the AKP from the energy levels in the QM side using 
\eqref{eq:height} derived from 
\eqref{eq:delta-like}. (For the half periodic orbits a suitable modification is 
necessary).

In Fig.~\ref{fig:us-exp_comparison} 
the horizontal axis is for $\lambda_{\Gamma_{\rm PO}}$
and the vertical axis is for $\lambda^*$. 
The points 
($\lambda_{\Gamma_{\rm PO}}$,
$\lambda^*$) lie almost along the line $y=x$ and
they come the closer to the line for the larger $j_{\rm max}$.
Provided that QM data with sufficiently large $j_{\rm max}$ 
should be  obtained it should be possible to extract an accurate
 information on each of the $\lambda_{\Gamma_{\rm PO}}$.
Considering the conceptual gap between 
the fluctuation in the energy levels 
(calculated from the diagonalization of Schr\"odinger equation) 
 and the Lyapunov stability exponents of classical periodic orbits,
this possibility is intriguing.

Summing up, sufficient information is embedded in the QM side which enables us to extract the actions and Lyapunov exponents
of the AKP periodic orbits, provided that
the POT side is properly formulated so that it accommodates
the symmetry in the QM side \cite{gutzwiller82}.

\subsection{The verification that the POT side correctly accommodates the symmetries of the QM side.}

\begin{figure}[!bp]
\begin{center}
  \includegraphics[width=110mm,clip]{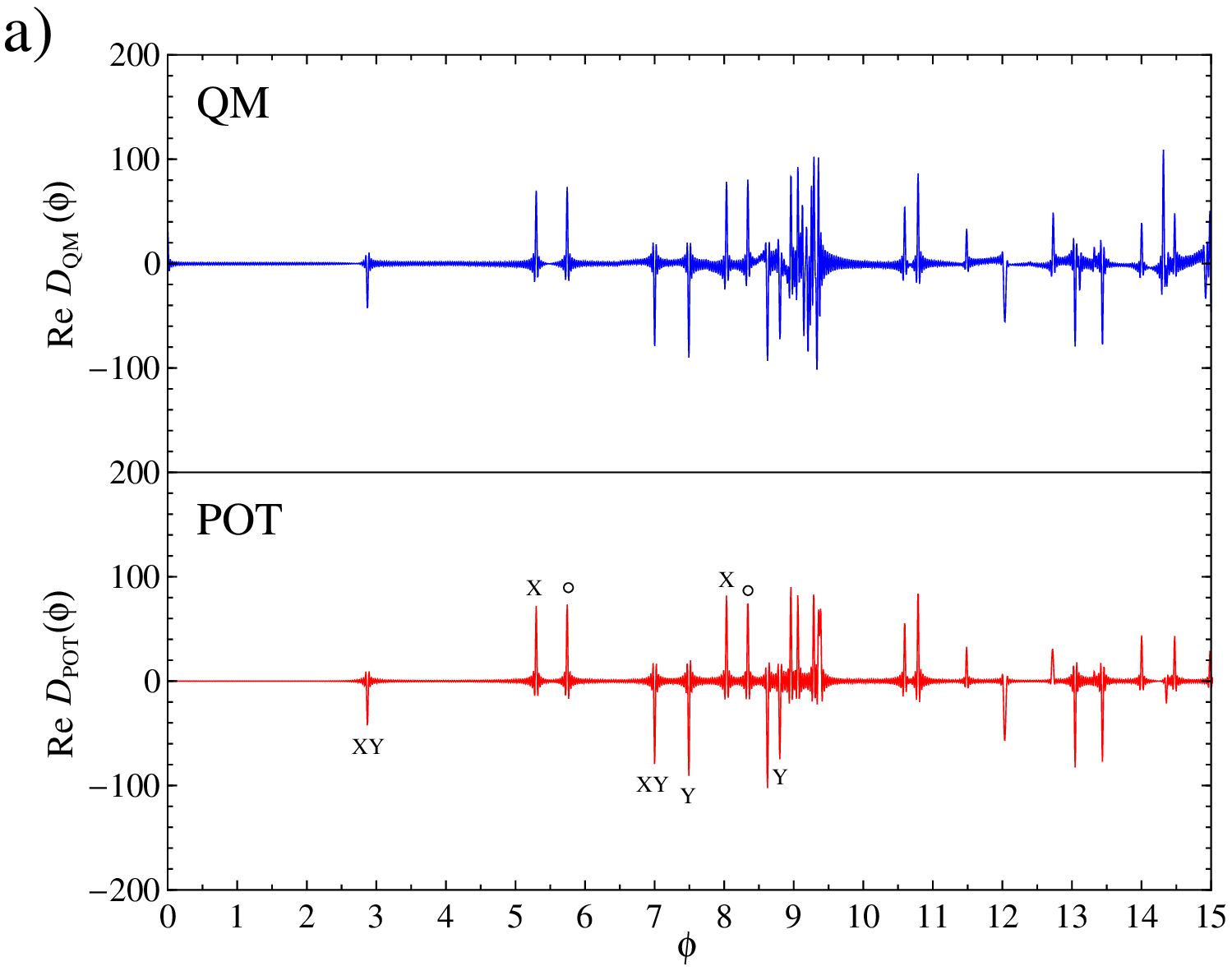}
\end{center}
\begin{center}
  \includegraphics[width=110mm,clip]{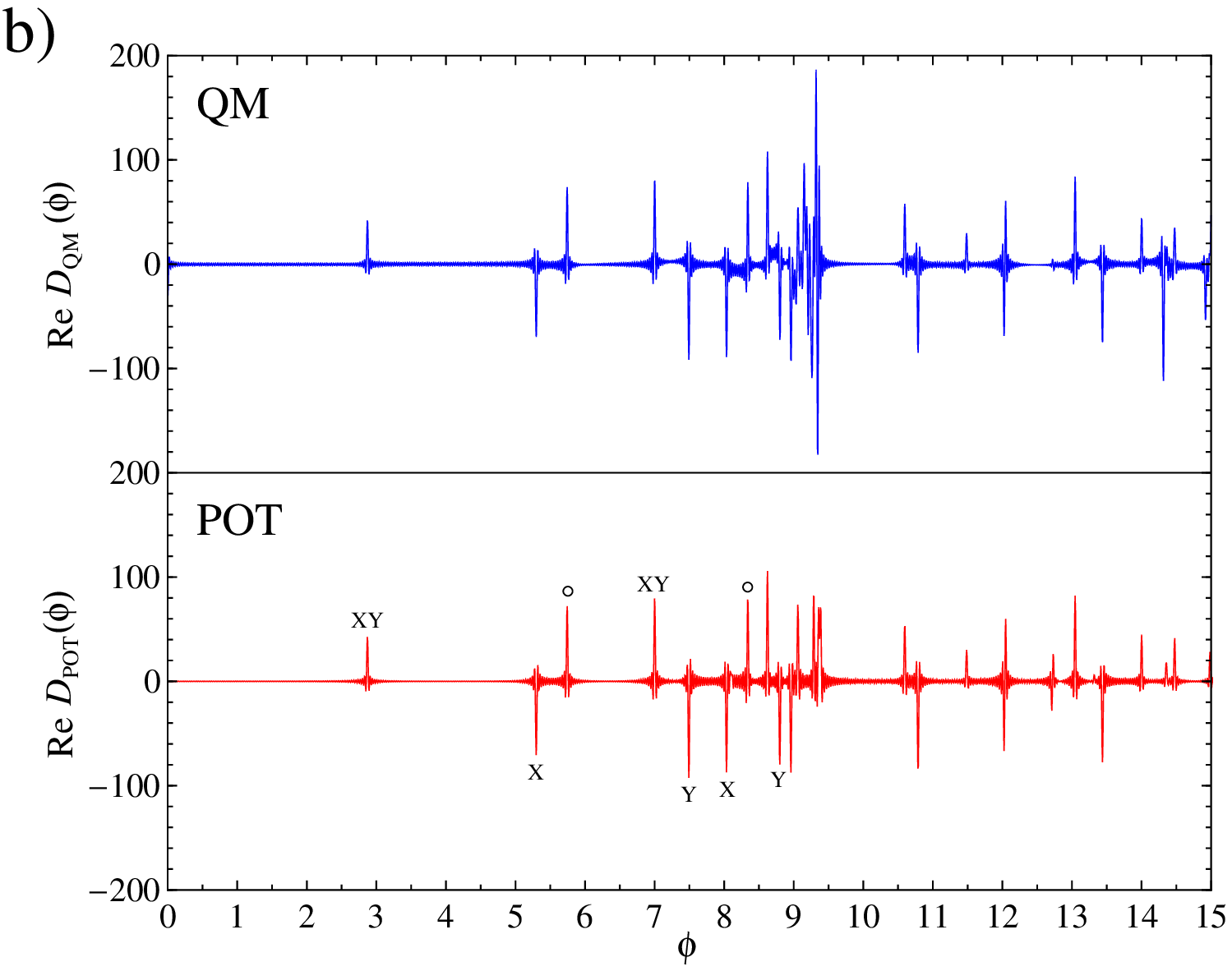}
\vspace{-5mm}
\end{center}
\caption{a) parity even and $m=0$ sector with $j_{\rm max}=10^4$, 
b) parity odd and $m=0$ sector with $k_{\rm max}=10^4$.
Real parts of $D_{\rm QM}(\phi)$ (upper) and 
$D_{\rm POT}(\phi)$ (lower) are compared for each sector.
The peaks in Re$D_{\rm POT}(\phi)$
of full, $X$-, $Y$-, and $XY$- symmetric POs
are signified by symbols--circle, X, Y, XY respectively; 
they appear, according to trace formulas,
$(+,+,-,-$) for parity even and ($+,-, -,+$) for parity odd.
The peaks in Re$D_{\rm QM}(\phi)$ follows perfectly this sign rule
and all the heights agree with the POT prediction.
The symmetries in the QM side is properly accommodated in the POT orbifolding
formulation. For the accumulation region ${\mathcal R}~[\phi\in (8.5,9.5)]$, 
inclusion of PO with $2N \ge 14$
will improve the POT prediction. However we have been unable to do it  because shooting for those are difficult 
and also because the Gutzwiller's 
approximation formula \cite{gutzwiller80, gutzwiller81, gutzwiller82} 
becomes inaccurate for the actions of particular POs 
(those evolving along the heavy $x$-axis) contributing to ${\mathcal R}$.
}
\label{fig:amp_comparison}
\end{figure}

Up to now we have compared 
$D_{\rm QM}(\phi)$ and $D_{\rm POT}(\phi)$
after taking the absolute value squared.
Now let us compare 
in Fig. \ref{fig:amp_comparison}  
the real parts of 
$D_{\rm QM}(\phi)$
with   
$D_{\rm POT}(\phi)$
for the even and odd parity states
separately.
The latter $D_{\rm POT}(\phi)$ 
is calculated by Fourier-Laplace formation 
from  the symmetrized response functions  
$g_e^{m=0}(E)$ and $g_o^{m=0}(E)$
in \eqref{eq:full-resp_fn}. 
Therefore, for this comparison, the rule of sign 
factors in the POT side is essential 
and we can test whether the 
POT side under the symmetrization in terms of
the full and half orbits correctly accommodate 
the symmetries ($m=0$
and parity $\pm$) of the QM side.
The rule consists of three steps.
(We call the peak with the positive height as 
`$+$ peak' and that with negative height as 
`$-$ peak').

(i) It is easy to see from
\eqref{eq:trace_formula_X},
\eqref{eq:trace_formula_Y},
\eqref{eq:trace_formula_XY}
and
\eqref{eq:trace_formula_akp},
that the relative sign of the imaginary part  
of 
$g_X(E)$, $g_Y(E)$ and $g_{XY}(E)$
to $g(E)$ is  ($+,-,-$) respectively.

(ii) We see that the $D_{\rm POT}$ in \eqref{eq:FT_cut}
derived from $g(E)$ by Fourier-Laplace transformation 
\eqref{eq:fourier-cut}
has a positive peak at every $\phi=\Phi_{\Gamma_{\rm PO}}$.
Combining this with (i), the sign of every peak
in 
$D(\phi)$,
$D_{X}(\phi)$,
$D_{Y}(\phi)$,
$D_{XY}(\phi)$
derived 
from $g_X(E)$, $g_Y(E)$ and $g_{XY}(E)$
are ($+,+,-,-$) 
respectively.

(iii) Finally, 
$D_e^{m=0}(\phi)$ and $D_o^{m=0}(\phi)$
are respectively given by
\begin{align} 
\label{eq:D_fn}
\nonumber
D_e^{m=0}(\phi)
&=\frac{1}{4}\left[ D(\phi)+D_Y(\phi)+D_X(\phi)+D_{XY}(\phi)\right]\\
D_o^{m=0}(\phi)
&=\frac{1}{4}\left[ D(\phi)+D_Y(\phi)-D_X(\phi)-D_{XY}(\phi)\right]
\end{align}
corresponding to \eqref{eq:full-resp_fn}.
Therefore, combining with (ii), we see that
the full- , $X$-, $Y$-, and $XY$-symmetric orbit
respectively predict ($+,+,-,-$)-peaks for the parity even sector,
while they predict ($+,-, -,+$)-peaks for the parity odd sector.

The comparison in Fig. \ref{fig:amp_comparison}
shows that the peaks (and the background) 
produced by the sum of 
$\exp(i u_j \phi)$ in the QM side 
perfectly agree 
with those predicted by the POT side
with respect to the heights, widths as well as
the signs, and also for both the 
parity even and odd sectors.
This succinctly shows that 
POT side successfully accommodates 
the symmetry of the sectors in the QM side.

\section{Conclusion}
\label{sec:conclusion}
In this paper we have shown how accurately Gutzwiller's trace 
formula catches the quantum physics of AKP in the completely ergodic region
$\mu=1/\nu=\sqrt{5}$ (or equivalently $\gamma=0.2$).
This in turn has lead us to the observation that the information of classical chaotic POs is 
amazingly embedded in QM side of AKP; 
we can extract it by our formulation in which the classical periodic orbits
make peaks with minimal interference in the POT side. 

The trace formula is based on the Feynman's path integral 
formula \eqref{eq:prop} for the propagator,
and we have called the left and right hand sides
QM and POT sides respectively. 
After briefly introducing the trace formula, we followed Gutzwiller's idea \cite{gutzwiller82}
and gave trace formulas for the quantum subspaces with definite symmetries.

Restriction to the parity even (odd) sector in three dimensional AKP quantum physics
can be accounted by an introduction of a half periodic orbit
$\Gamma_X$ which closes by $X$-transformation (periodic in ${\mathcal R}^2/X$, 
see Fig.~\ref{fig:x_po})
and taking the symmetrization (anti-symmetrization) according to
\eqref{eq:resp_fn_parity_separated}.  
We remarked there that the response function $g(E)$ (in the POT approximation) 
has 
$\sinh(\lambda_{\Gamma_{\rm PO}}/2)$ 
in the denominator from the contribution of each full periodic orbit $\Gamma_{\rm PO}$, 
while $g_X(E)$ has 
$\cosh(\lambda_{\Gamma_{\rm X}}/2)$ 
from the contribution of each $\Gamma_X$.
See \eqref{eq:trace_formula_akp} and \eqref{eq:trace_formula_X}.
Furthermore, the restriction to the quantum sector with magnetic quantum number $m=0$ 
is realized in the POT side by the introduction of $g_{\rm Y}(E)$.

The trace formula is usually tested by predicting 
the energy levels  (or d.o.s.) from the POT side.
Here the prediction is given under the interference of  
amplitudes of periodic orbits and hence, in this direction,
it is impossible to separately investigate periodic orbits one by one.  

Our trace formulas (\eqref{eq:FT_full} and its symmetry partners)
are basically a Fourier transformation of usual ones 
but with a one shot idea of using the scaling property of the AKP 
action by the introduction of a convenient variable $u=1/\sqrt{-2 E}$.

The new test of quantum-classical correspondence is done in
Fig.~\ref{fig:ft3d} and Fig.~\ref{fig:amp_comparison},
and 
we have shown that the information of periodic orbits 
can be indeed extracted from the quantum energy levels 
in Fig.~\ref{fig:s_comparison} and Fig.~\ref{fig:us-exp_comparison}.

The QM side of \eqref{eq:FT_cut} is a sum of the sequence $\exp{(u_j\phi)}$ 
where $u_j$ is determined from the energy levels --- the eigenvalues of the 
Schr\"odinger equation. 
We have worked out the energy levels with sweat,
but, by looking at them,  it is hardly ever understood how the interference between $\exp{(u_j\phi)}$ 
makes up the peaks at the prescribed position by the periodic orbits, which are the
solution of classical Hamiltonian equation of motion.
In particular, the fluctuation of the $E_j$ is essential for this make up;
if we substitute Thomas-Fermi approximation for the levels (the averaged $E_j$)
the peaks in $D_{\rm QM}(\phi)$ disappear instantly.
It is remarkable that the intricate fluctuation can produce via the inverse trace formula the POs 
that are smooth solutions of EOM. 
One interesting challenge in the inverse quantum chaology may be the test whether, in the diamagnetic Kepler problem,
it extracts from quantum energy levels the salient distinction between the Lyapunov exponents
of the POs that are longitudinal and transverse to the field direction \cite{wintgen87lyapunov}.

At the anisotropy $\mu=1/\nu=\sqrt{5}$ (or $\gamma=0.2$), 
the classical phase space is filled by isolated unstable periodic orbits. We have motivated this work 
as the first step towards the most ambitious study of the intermediate region 
around ($\mu \simeq 1.1$, $\gamma \simeq 0.8$) between chaos and regularity, 
where the classical phase space is deemed to be dominated by the web of chaos 
\cite{verbaarschot03, garcia08}.  
We have already calculated the $10^4$ energy levels at every $\gamma$ 
in the interval $\gamma \in (0.2, 0.85)$ and inclement $0.005$ for the QM side.
On the other hand we have observed that POs gradually disappear with the increase of $\gamma$. 
We are going to investigate how the trace formula \eqref{eq:FT_full} reflects this situation. 
The intermediate region of AKP is a fascinating region in that it is conjectured
that the wave functions are multifractal under deep connection to the Anderson localization
\cite{garcia08,verbaarschot03,ourpaper12,ourpaper13}.  
We will challenge this region in a forthcoming paper. 

\section*{Acknowledgement}
We would like to take this opportunity to thank Professor M. Gutzwiller
for answering our question in 2008 in regarding the reference \cite{gutzwiller77} and 
kindly letting us know of the reference \cite{gutzwiller81}, which encouraged us 
very much. 
We thank M. Hentschel, S. Mochizuki Nishigaki, T. Takami, I. Tsutsui, K. Sumiya, H. Uchiyama
for useful discussions, and Meiji University for the priority research fund.  
One of us (K.K.) thanks Martina Hentschel for her kind hospitality in Germany,
and his new colleagues at Max-Planck-Institut f\"ur Physik Komplexer Systeme, Germany.


\begin{thebibliography}{99}
\bibitem{einstein17} 
A. Einstein, 
Zum Quantensatz von Sommerfeld und Epstein,
Verh. Dtsch. Phys. Ges. {\bf 19}, 82-92 (1917).

\bibitem{brillouin26} 
L. Brillouin,
Remarques sur la mecanique ondulatoire (Notes on undulatory mechanics),
J. phys. radium 7, 353-368 (1926).

\bibitem{keller58} 
J. B. Keller, 
Corrected Bohr-Sommerfeld Quantum Conditions for Nonseparable Systems,
Ann. Phys. (NY) {\bf 4}, 180-188, (1958).

\bibitem{dstone05}
 A. Douglas Stone,
Einstein's unknown insight and the problem of quantizing chaos, 
Physics Today {\bf  58}, 8, p. 37-43 (August, 2005).

\bibitem{marcus92}
C. M. Marcus et al., 
Conductance fluctuations and chaotic scattering in ballistic microstructures,
Phys. Rev. Lett. \textbf{69}, 506-509 (1992).

\bibitem{weiss91}
D. Weiss {\it et al.}, 
Electron pinball and commensurate orbits in a periodic array of scatterers,
Phys. Rev. Lett. \textbf{66}, 2790-2793 (1991).

\bibitem{chaudhury09}
S. Chaudhury, A. Smith, B. E. Anderson, S. Ghose, P. S. Jessen,
Quantum signatures of chaos in a kicked top,
Nature \textbf{461}, 768 (2009).

\bibitem{stein92}
J. Stein, H. -J. St\"ockmann,
Experimental determination of billiard wave functions,
Phys. Rev. Lett. \textbf{68}, 2867-2870 (1992).

\bibitem{heller84}
E. J. Heller,
Bound-state eigenfunctions of classically chaotic hamiltonian systems:
scars of periodic orbits,
Phys. Rev. Lett. \textbf{53}, 1515-1518 (1984).

\bibitem{heller89}
E. J. Heller, 
Wavepacket dynamics and quantum chaology,
{\it in} Giannoni, M. J., Voros A. \& Zinn-Justin J. (ed.)
{\it Les Houches Summer School, Session LII, Chaos and quantum physics},
North-Holland, Amsterdam, 547-664 (1989).

\bibitem{zhou09}
Z. Chen {\it et al.}, 
Realization of anisotropic diamagnetic Kepler problem 
in a solid state environment, 
Phys. Rev. Lett. \textbf{102}, 244103, pp.1-4 (2009).

\bibitem{zhou10}
W. Zhou {\it et al.}, 
Magnetic field control of the quantum chaotic dynamics of 
hydrogen analogs in an anisotropic crystal field, 
Phys. Rev. Lett. \textbf{105}, 024101, pp.1-4 (2010).

\bibitem{hasan10}
M. Z. Hasan, C. L. Kane,
Colloquium: Topological insulators,
Rev. Mod. Phys. \textbf{82}, 3045-3067 (2010). 

\bibitem{kitaev09}
A. Kitaev,   
Periodic table for topological insulators and superconductors,
AIP Conf. Proc. \textbf{1134}, 22-26 (2009).

\bibitem{ryu10}
S. Ryu, A. Schnyder, A. Furusaki, and A. W. W. Ludwig, 
Topological insulators and superconductors: tenfold
way and dimensional hierarchy,
New J. Phys. \textbf{12}, 065010 (2010).

\bibitem{gutzwiller71}
M. C. Gutzwiller, 
Periodic orbits and classical quantization conditions, 
J. Math. Phys. \textbf{12}, 343-358 (1971).


\bibitem{miller75}
W. H. Miller, 
Semiclassical quantization of nonseparable systems:
A new look at periodic orbit theory, 
J. Chem. Phys. \textbf{75}, 996-999 (1975).

\bibitem{gutzwiller80}
M. C. Gutzwiller, 
Classical quantization of a Hamiltonian with ergodic behavior, 
Phys. Rev. Lett. \textbf{45}, 150-153 (1980).

\bibitem{gutzwiller82}
M. C. Gutzwiller, 
The quantization of a classically ergodic system,
Physica D \textbf{5}, 183-207 (1982).

\bibitem{yellowbook} 
M. C. Gutzwiller, 
{\it Chaos in Classical and Quantum Mechanics}, 
Springer (1990).

\bibitem{wintgen88b}
D. Wintgen, 
Semiclassical path-integral quantization of nonintegrable Hamiltonian systems, 
Phys. Rev. Lett. \textbf{61}, 1803-1806 (1988).


\bibitem{wigner51}
E. P. Wigner, 
On the statistical distribution of the widths and spacings of nuclear resonance levels,
{\it Proc. Cambridge Philos. Soc.} \textbf{47}, 790-798 (1951).

\bibitem{dyson62a}
F. J. Dyson, 
Statistical theory of the energy levels of complex systems. I, II, III,
J. Math. Phys. \textbf{3}, 140-156, 157-165, 166-175 (1962).

\bibitem{dyson62b}
F. J. Dyson, 
The threefold way. Algebraic structureof symmetry groups
and ensembles in quantum mechanics,
J. Math. Phys. \textbf{3}, 1199-1215 (1962).

%

\bibitem{altland97}
A. Altland and M. Zirnbauer,
Nonstandard symmetry classes in mesoscopic normal-superconducting hybrid structures,
Phys. Rev. B55, 1142 (1997).

\bibitem{bgs84}
O. Bohigas, M. J. Giannoni, and C. Schmit, 
Characterization of chaotic quantum syectra and universality of level fluctuation laws,
Phys. Rev. Lett. \textbf{52}, 1 (1984).


\bibitem{sieber01}
M. Sieber and K. Richter, 
Correlations between periodic orbits and their r\^ole in spectral statistics,
Physica Scripta \textbf{T 90}, 128 (2001).

\bibitem{sieber02}
M. Sieber, 
Leading off-diagonal approximation for the spectral form factor for uniformly hyperbolic systems, 
J. Phys. A \textbf{35}, L613 (2002)

\bibitem{heusler07}
S. Heusler, S. M{\"u}ller,  A. Altland, P. Braun, and F. Haake,
Periodic-orbit theory of level correlations
Phys. Rev. Lett.  \textbf{98}, 044103, pp1-4 (2007).


\bibitem{gutzwiller77}
M. C. Gutzwiller, 
Bernoulli sequences and trajectories in the anisotropic Kepler problem, 
J. Math. Phys. \textbf{18}, 806-823 (1977). 

\bibitem{gutzwiller81}
M. C. Gutzwiller, 
Periodic orbits in the anisotropic Kepler problem, 
in {\it Classical Mechanics and Dynamical systems,}
edited by R. L. Devaney and Z. H. Nitecki, 
69-90 (1981),
Marcel Dekker, New York.

\bibitem{chazarain74}
J. Chazarain,
Formule de Poisson pour les vari\'et\'es riemanniennes,
Invent. Math. \textbf{24}, 65-82 (1974).\\
J. Chazarain, Spectre D'Un hamiltonien quantique et mecanique classique, 
Commun. Part. Diff. Eq., \textbf{5:6}, 595-644 (1980).

\bibitem{berry-speech}
M. Berry, {\it Gutzwiller 70th birthday in Some speeches and introductions},

http://www.phy.bris.ac.uk/people/berry\_mv/the\_papers/Speeches.pdf

\bibitem{wintgen87}
D. Wintgen, 
Connection between long range correlations in quantum spectra and classical periodic orbits,
Phys. Rev. Lett. \textbf{58}, 1589 (1987).

\bibitem{wintgen88a}
D. Wintgen and H. Marxer, 
Level statistics of a quantized cantori system, 
Phys. Rev. Lett. \textbf{60}, 971-974 (1988).

\bibitem{verbaarschot03}
A. M. Garc\'ia-Garc\'ia and J. J. M. Verbaarschot, 
Critical statistics in quantum chaos and Calogero-Sutherland model at finite temperature, 
Phys. Rev. \textbf{E 67}, 046104, pp.1-13 (2003).

\bibitem{zaslavsky91}
G. M. Zaslavsky,
Stochastic webs and their applications,
Chaos \textbf{1}, 1-12 (1991).

%


\bibitem{ourpaper12}
K. Kubo and T. Shimada, 
Anisotropic Kepler problem and critical level statistics, 
chapter 5 of {\it Theoretical Concepts of Quantum Mechanics}, 81-106 (2012), InTech. 

\bibitem{ourpaper13}
K. Sumiya, H. Uchiyama, K. Kubo, and T. Shimada, 
Classical and quantum correspondence in anisotropic Kepler problem, 
chapter 2 of {\it Advances in Quantum Mechanics}, 23-40 (2013), InTech.

\bibitem{devaney78}
R. L. Devaney, 
Collision orbits in the anisotropic Kepler problem, 
Inventiones. Math. \textbf{45}, 221-251 (1978).

\bibitem{ourpaper13a}
K. Sumiya,  K. Kubo, and T. Shimada, 
in preparation.

\bibitem{faulkner69}
R. A. Faulkner, 
Higher donor excited states for prolate-spheroid conduction bands: A reevaluation of silicon and germanium, 
Phys. Rev. \textbf{184}, 713-721 (1969).

\bibitem{tanner91}
G. Tanner, P. Scherer, E. B. Bogomonly, B. Eckhardt  and D. Wintgen,
Quantum eigenvalues from classical periodic orbits,
Phys. Rev. Lett.  \textbf{67}, 2410-2413 (1991).

\bibitem{wintgen94}
K. M\"uller and D. Wintgen, 
Scars in wavefunctions of the diamagnetic Kepler problem, 
J. Phys. B: At. Mol. Opt. Phys. \textbf{27}, 2693-2718 (1994).

\bibitem{wintgen87a}
D. Wintgen, H. Marxer, and J. S. Briggs, 
Efficient quantisation scheme for the anisotropic Kepler problem, 
J. Phys. \textbf{A 20}, L965-L968 (1987).

\bibitem{robbins89}
J. M. Robbins, 
Discrete symmetries in periodic-orbit theory, 
Phys. Rev. \textbf{A 40}, 2128-2136 (1989).

\bibitem{wintgen87lyapunov}
D. Wintgen, 
Calculation of Lyapunov exponent for periodic orbits
of the magnetised hydrogen atom,
J. Phys. B, At. Mol. Phys. \textbf{20}, L511-L515 (1987).

\bibitem{garcia08}
A. M. Garc\'ia-Garc\'ia and J. Wang, 
Universality in quantum chaos and the one-parameter scaling theory,
Phys. Rev. Lette \textbf{100}, 070603, pp.1-4 (2008).
\end{thebibliography}
\end{document}